# BEAM PHASE SPACE TOMOGRAPHY AT FAST ELECTRON LINAC AT FERMILAB*

A. Romanov[†], Fermilab, [60510] Batavia, USA

*Abstract*

FAST linear accelerator has been commissioned in 2017. Experimental program of the facility requires high quality beams with well-defined properties. Solenoidal fields at photoinjector, laser spot shape, space charge forces and other effects can distort beam distribution and introduce coupling. This work presents results of a beam phase space tomography for a coupled 4D case. Beam was rotated in two planes with seven quads by 180 degrees and images from YaG screen were used to perform SVD based reconstruction of the beam distribution.

## INTRODUCTION

Known distribution of the beam at some point in full 4D phase space can greatly improve understanding of the processes in the upstream section of the linear accelerator. Beam phase space tomography is a common name for several techniques that can use a set of beam projections for various settings in the upstream section to reconstruct beam distribution in a phase space with higher number of dimensions. Typically, problem is solved for one plane only, assuming no coupling with other degrees of freedom [1,2,3].

Another disadvantage of the most available methods is presence of the high-frequency artefacts in the reconstructed phase space that stretches far away from the beam's core [2,3]. In many cases it is a result of the amplified noise and systematic distortions in the experimental data. Presence of this errors can mask halo details and especially harmful for direct calculation of the beam's second moments matrix.

Presented technique is potentially free from both described problems, but requires a lot of computational power to reconstruct beam phase space distribution in fine details.

## INVERSE PROBLEM SOLVER

The success in solving inverse problems using Singular Value Decomposition (SVD) of linearized model, inspired its use in the beam phase space tomography method. Given a set of experimental data $V_{exp,j}$ is available the goal is to find the parameters $P_i$ of the model $\mathcal{M}_j(P_i)$ that best describes the measurements:

$$V_{mod,j} = \mathcal{M}_j(P_i)\, s_j. \quad (1)$$

Here $s_j$ are data weights, normally they are inversely proportional to the errors $\epsilon_j$ of the experimental data. In the case of a liner model, the task simplifies to:

$$\Delta V_{mod,j} = \mathcal{M}_{ji}\, s_j k_i \frac{\Delta P_i}{k_i}, \quad (2)$$

___________
* Work supported by the DOE contract No. DEAC02-07CH11359 to the Fermi Research Alliance LLC.
† aromanov@fnal.gov

here $k_i$ are the normalization coefficients that can be used to normalize influence of the physically different parameters.

The difference between the experimental data and the model is:

$$D_j = (V_{exp,j} - V_{mod,j}). \quad (3)$$

The goal is to find a variation of the parameters $P_i$ that cancels the difference between model and experimental data:

$$\Delta V_{mod,j} = -\Delta D_j = D_j. \quad (4)$$

The model parameters variation can be obtained from here by applying pseudo inversion to the $(\mathcal{M}_{ji}\, s_j k_i)$. Singular Values Decomposition (SVD) is a powerful method for such calculation. One remarkable feature of this technique is easy control over the influence of statistical errors in the experimental data on the output result. Application of SVD gives the parameters variation as:

$$\Delta P_i = k_i (\mathcal{M}_{ji}\, s_j k_i)^{-1}_{SVD} D_j. \quad (5)$$

For the nonlinear models Jacobean matrix should be calculated from the model and several iterations are necessary to converge to a solution.

## PHASE SPACE TOMOGRAPHY

*Phase space description and projections*

The idea of the proposed method is to represent beam phase space distribution by a set of voxels in multidimensional space with model parameters $P_i$ as their intensities. The experimental data $V_{exp,j}$ is composed of all informative pixels from all the projections measured by a selected camera. Known optics model allows to transport the phase space voxels to the location of the screen and calculate X-Y projections.

To optimize voxel distribution, it is useful to define latter in a normalized phase space. If initial guess on the values of the second-moments matrix $M_0$ is known, then eigenvalues of matrix $(i\, S\, M_0)$ are mode RMS emittances of the beam $(\epsilon_1, \epsilon_1, \epsilon_2, \epsilon_2)$. Here S is fundamental symplectic matrix. Eigenvectors $Y_k$ and $Y_k^*$ of the same matrix $(i\, S\, M_0)$ give the basis of the phase space, and can be used to construct a diagonalization matrix for the arbitrary second moment matrix [4]:

$$T_{diag}(Y_1,\dots,Y_n,\psi_1,\dots,\psi_n) =$$
$$(Re[Y_1\, e^{\psi_1}], Im[Y_1\, e^{\psi_1}],\dots, Re[Y_n\, e^{\psi_n}], Im[Y_n\, e^{\psi_n}]). \quad (6)$$

For the eigenvector normalization:

$$Y_n^t S Y_k^* = -\frac{2i\delta_{nk}}{\epsilon_k}, \quad (7)$$

second moment matrix $M_0$ will become unit after transformation into normalized phase space:

$$T_{diag}^t\, M_0 T_{diag} = I. \quad (8)$$

Phase space distribution of the beam was represented by a set of 4D voxels distributed in a regular pattern along the

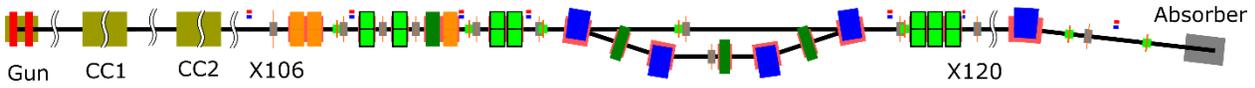
Figure 1: The FAST electron linac: low energy beamline was commissioned to the beam energies up to 52.5 MeV [6] and the high energy transport section is now operational up to 300 MeV

cartesian grid in a beam's phase space normalized by its initial conditions. Each voxel added the same distribution scaled by its amplitude and shifted to the corresponding 4D point. A multivariate normal distribution was selected for the voxel's distribution function, because it has trivial projections on the sub-spaces, such as 2D X-Y images. Covariance matrix of the voxel's distribution function in a normalized phase space at base point was a unit matrix multiplied by scaling coefficient $K/n_{steps}$. Here $n_{steps}$ is a grid dimension in one direction and K should be picked around 0.5 to allow uniform coverage of the space. Extent of the grid in normalized phase space at base point was selected to be bigger than 1 to catch details of the halo.

*Lattice configuration*

Cross X106 was selected as a reference point for the presented phase space analysis (Figure 1). Screen at the cross X120 that is located near the end of the low energy section was used to record a series of 2D images of the beam. Seven of the eight available quadrupoles allowed to vary phase advance in one plane while keeping constant beam Twiss parameters ($\beta_x, \alpha_x, \beta_y, \alpha_y$) and phase advance in the other plane. This approach allows to keep beam transverse shape roughly constant while rotating beam by $\pi$ in steps of $\pi/10$, resulting in total of 22 projections. Mentioned phase scan lattices setup depends on the specific initial conditions and was done once for the typical anticipated beam parameters at the reference point. Figure 2 and Figure 3 show beta functions for the 11 lattices for the rotation in X plane.

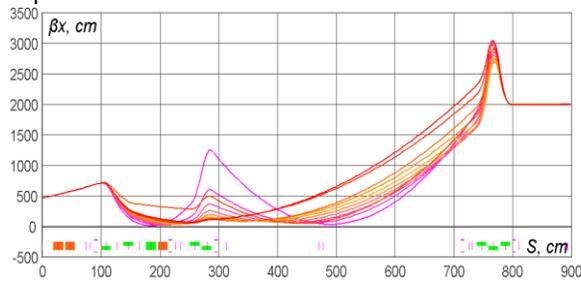
Figure 2: $\beta_x$ for all lattices in scan X

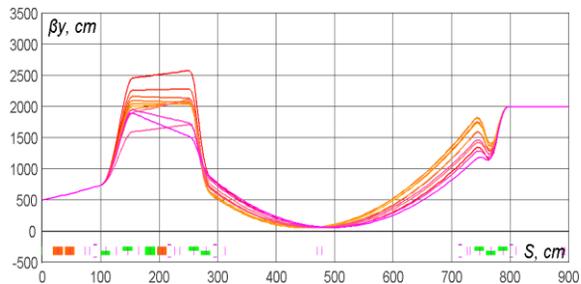
Figure 3: $\beta_y$ for all lattices in scan X

*Initial conditions*

In order to better match multidimensional greed to the phase space volume occupied by the beam, RMS initial conditions of the latter must be found prior to a tomography. Second moments of the beam from the same set of images were extracted. Initial Twiss parameters and emittances was fitted to match model predictions and experimental data using described SVD-based solver, see fitted data on the Figure 4. Initial conditions were intentionally restricted to the uncoupled case, to see if the tomography can reveal small but present correlation terms.

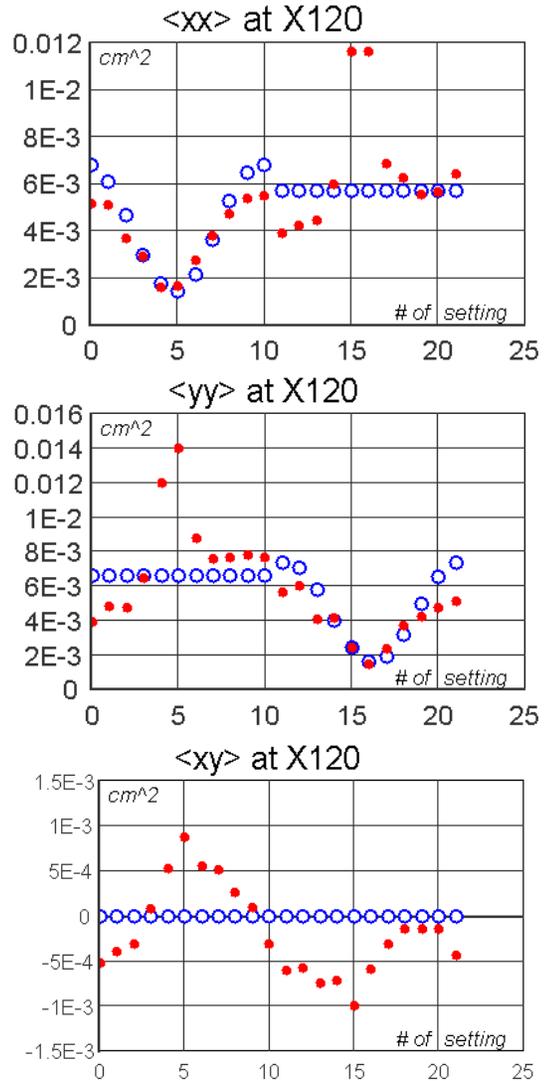
Figure 4: Measured second moments of the beam for all 22 scan lattice configurations (red circles) with best fit for decoupled initial conditions (blue circles)

*Results*

Presented are results for 32 MeV, 100 nC bunches. Each of 22 images was fitted by 2D gaussian distribution, centered, trimmed to ($\pm 3\sqrt{<xx>}$, $\pm 3\sqrt{<yy>}$), and scaled down by x10. This resulted in a total of 40659 data points in the experimental data set.

Phase space was filled with a 6x6x6x6 points resulting in 1296 model variables. Singular value decomposition took ~1 hour of computation on a single core of Intel Xeon CPU E5-1650 3.5GHz. Because of the linear nature of the solved task found matrices can be reused for the same scanning procedure if beam initial distribution stays within used grid.

Figure 5 shows sorted singular values (SV) of the $\mathcal{M}_{ji}$ in log scale, with magenta rectangle around selected SVs. Pseudoinverse matrix was calculated using 200 singular values.

One of the criteria for truncation of SV spectrum is absence of negative intensities for the voxel amplitudes. Iterative methods can be used to avoid negative intensities.

Projections of the reconstructed phase space are presented in Figure 6. Figure 7 and Figure 8 show measured profiles for scans in X and Y planes along with corresponding projections of found model.

## SUMMARY

Presented method demonstrated good potential for reconstruction of the artefact-free fully coupled 4D phase space distributions. To further scale this method towards reconstruction of the distributions with higher detailing or in 5 dimensions optimized matrix manipulation algorithms should be implemented. Resulting $\mathcal{M}_{ji}$ has a very big size, but only small fraction has non-zero values. In addition, $\left(\mathcal{M}_{ji} s_j k_i\right)^{-1}_{SVD}$ was calculated basing on 20% of the singular values. Both of these features make logical to use specialized packages optimized for sparse arrays and partial SVD, which is the next step planned by author.

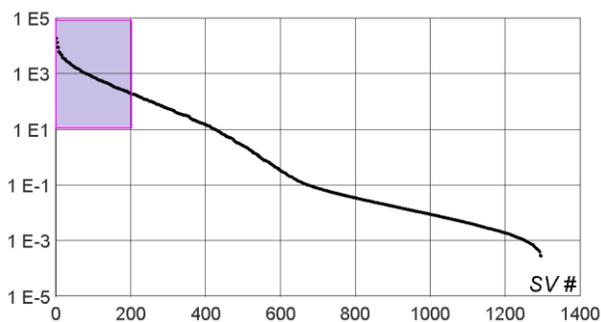

Figure 5: SV spectra with selected values

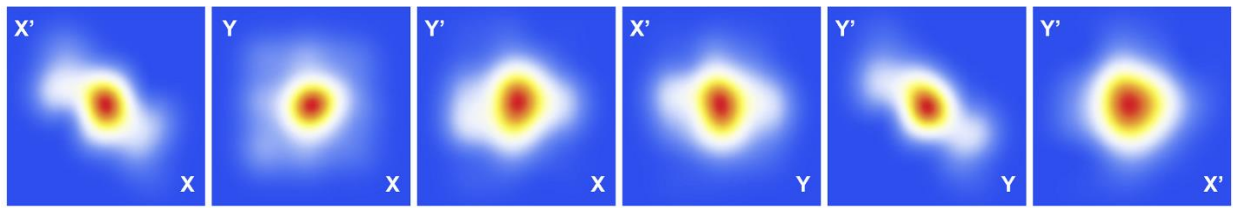

Figure 6: 2D projections of the reconstructed 4D distribution

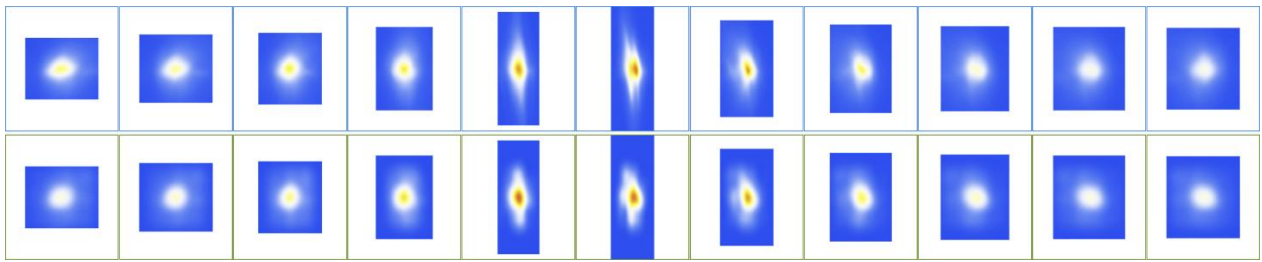

Figure 7: Prepared images and corresponding projections from reconstructed 4D distribution for scan in X plane

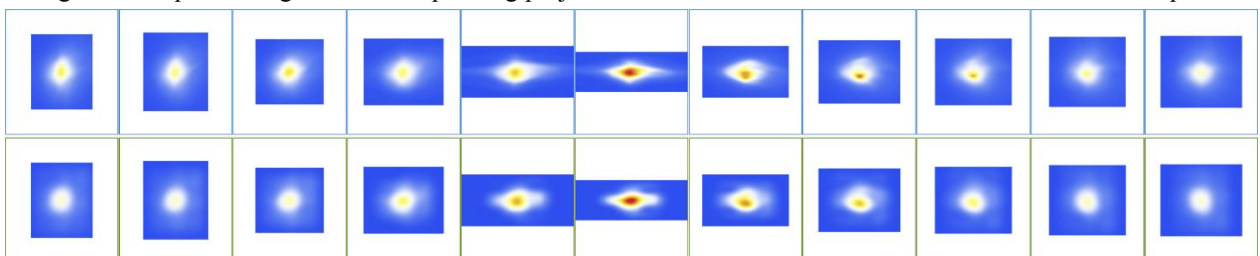

Figure 8: Prepared images and corresponding projections from reconstructed 4D distribution for scan in Y plane